\documentclass[twocolumn,floatfix,pre]{revtex4-1}
\usepackage{graphicx}
\usepackage[]{tikz} 
\usetikzlibrary[arrows]
\usepackage{amsmath}
\usepackage{bm}
\usepackage[caption=false]{subfig}
\begin{document}
\title{Stochastic thermodynamics  for a periodically driven single-particle pump}
\author{Alexandre Rosas}
\email{arosas@fisica.ufpb.br}
\affiliation{Departamento de F\'{\i}sica, CCEN, Universidade Federal da Para\'{\i}ba, Caixa Postal 5008, 58059-900, Jo\~ao Pessoa, Brazil.}
\author{Christian Van den Broeck}
\affiliation{Hassel University, B-3590 Diepenbeek, Belgium.}
\author{Katja Lindenberg}
\affiliation{Department of Chemistry and Biochemistry, and BioCircuits Institute, University of California San Diego, La Jolla, California 92093-0340, USA.}

\begin{abstract}
We present the stochastic thermodynamic analysis of a time-periodic single particle pump, including explicit results for flux, thermodynamic force, entropy production, work, heat and efficiency. These results are valid far from equilibrium. The deviations from the linear (Onsager) regime are discussed.
\end{abstract}

\maketitle

\section{Introduction}
\label{sec:introduction}

Engines are to engineering what catalysts are to chemistry: they facilitate a transformation. Engines usually deal with the transformation between different forms  of energy,  while catalysts deal with the transformation of chemical substances. Since catalysts and engines are left untouched at the end of each completed transformation, and the entire operation needs to be repetitive to continue the process, catalysts and engines typically operate in cyclic fashion. The most famous example of a cyclic engine is undoubtedly the Carnot engine. In addition to the case of Carnot-like engines, the focus in the literature on irreversible thermodynamics has been on  the operation of engines under steady state conditions. The key properties of engines in the linear regime in this case are captured by the famous Onsager coefficients.

The purpose of this paper is to introduce a periodically driven single-particle pump that illustrates two recent developments in the field: the derivation of Onsager coefficients for periodically driven machines \cite{schmiedl,izumida1,izumida2,esposito,izumida3,izumida4,brandner1,proesmans1,proesmans2,benenti,brandner2,proesmans3}, and the thermodynamic description of small scale systems based on stochastic thermodynamics \cite{rosas16,rosas16b,proesmann4,proesmann5}. 
Our model has an additional virtue: it is exactly solvable even far away from the linear regime.

Our model is arguably the simplest exactly solvable example of such a construction. It consists of a system that can switch between two different configurations, with only two possible states in each configuration: empty or occupied.  In order to have a pumping function, the system needs to be placed in contact with (at least) two (ideal) reservoirs. The configurations are such that, in the absence of  switching, the system reaches a full equilibrium state in one or the other configuration. The nonequilibrium driving consists of a modulation, piece-wise constant in time, between the configurations. This modulation affects the exchange rates with the reservoirs. There is now a current which reflects the two mechanisms that break basic symmetries: a spatial asymmetry, and the alternation between two configurations that tend toward two different equilibrium states.    
We present the full thermodynamic picture, including explicit analytic expressions for entropy production, thermodynamic force, work, heat, and efficiency.

\section{Single Particle Pump}
\label{sec:model}

A pump is a construction that transports a ``conserved'' quantity (such as a particle) from one location to another. For concreteness, we consider the transport of particles. Being particularly interested in the stochastic aspects of the problem, we focus on the extreme limit of a pump that manipulates particles one at a time. More precisely, we assume that the system that connects the two (or more) reservoirs between which the particles are pumped can hold at most one particle at a time. We refer to the two possible states of the system as ``occupied'' and ``empty'',  and denote the  probability that the system is occupied by $p$. 

\begin{figure}[htpb]
  \begin{center}
  \begin{tikzpicture}[scale=1, transform shape]
    \draw (0,0) -- (1,0) -- (1,2.45) -- (0,2.45);
    \draw (0,2.55) -- (1,2.55) -- (1,5) -- (0,5);
    \node (R1l) at (-0.6,1.2) {left reservoir 1};
    \node (R2l) at (-0.6,3.7) {left reservoir 2};
    \draw (5,0) -- (4,0) -- (4,2.45) -- (5,2.45);
    \draw (5,2.55) -- (4,2.55) -- (4,5) -- (5,5);
    \node (R1r) at (5.5,1.2) {right reservoir 1};
    \node (R2r) at (5.5,3.7) {right reservoir 2};
    \draw [blue,fill=blue] (2.5,0.5) circle (0.25cm) node (fill1) [above,yshift=0.40cm] {configuration 1};
    \draw [blue] (2.5,1.9) circle (0.25cm) node (empty1){};
    \draw [red,fill=red] (2.5,3.05) circle (0.25cm) node (fill2) [above,yshift=0.40cm] {configuration 2};
    \draw [red] (2.5,4.45) circle (0.25cm) node (empty2){};
   \draw[blue,->,>=stealth,very thick] ([yshift=0.1cm] R1l.east) to [bend left]  ([xshift=-0.2cm]empty1.north) node [above,xshift=-0.7cm, yshift=-0.05cm] {$\omega_{\ell 1}$};
        \draw[blue,->,>=stealth,very thick] ([xshift=-0.2cm,yshift=-0.4cm]fill1.south) to [out=270,in=270] (R1l.south east)  node [below,xshift=1.0cm, yshift=-0.8cm] {$\omega_{1 \ell}$};    
  \draw[blue,->,>=stealth,very thick] ([yshift=0.1cm] R1r.west) to [bend right]  ([xshift=0.2cm]empty1.north) node [above,xshift=0.6cm, yshift=-0.1cm] {$\omega_{r1}$};   
    \draw[blue,->,>=stealth,very thick] ([xshift=0.2cm,yshift=-0.4cm]fill1.south) to [out=270,in=270] (R1r.south west)  node [below,xshift=-0.8cm, yshift=-0.8cm] {$\omega_{1r}$};
    \draw[red,->,>=stealth,very thick] ([yshift=0.1cm] R2l.east) to [out=90, in=135]  ([xshift=-0.2cm]empty2.north) node [above,xshift=-0.8cm, yshift=0.15cm] {$\omega_{\ell  2}$};
    \draw[red,->,>=stealth,very thick] ([xshift=-0.2cm,yshift=-0.4cm]fill2.south) to [out=270,in=270] (R2l.south east)  node [below,xshift=0.7cm, yshift=0.0cm] {$\omega_{2\ell}$};
    \draw[red,->,>=stealth,very thick] ([yshift=0.1cm] R2r.west) to [out=90,in=45]  ([xshift=0.2cm]empty2.north) node [above,xshift=0.6cm, yshift=0.1cm] {$\omega_{r 2}$};
    \draw[red,->,>=stealth,very thick] ([xshift=0.2cm,yshift=-0.4cm]fill2.south) to [out=270,in=270] (R2r.south west)  node [below,xshift=-0.7cm, yshift=0.0cm] {$\omega_{2 r}$};
    \draw[draw=none] (6,5.5) -- (-1,5.5);

  \end{tikzpicture}
  \end{center}
  \caption{Schematic representation of a single particle pump. A two-state system 
  that can operate in two different configurations, $1$ and $2$, is connected to two reservoirs. In each configuration, transition rates between system and reservoirs obey detailed balance even though the couplings may be asymmetric (that is, the couplings of the system to left and right reservoirs may be different). Pumping is achieved by periodic alternation from one configuration to the other.}
  \label{fig:model}
  \end{figure}
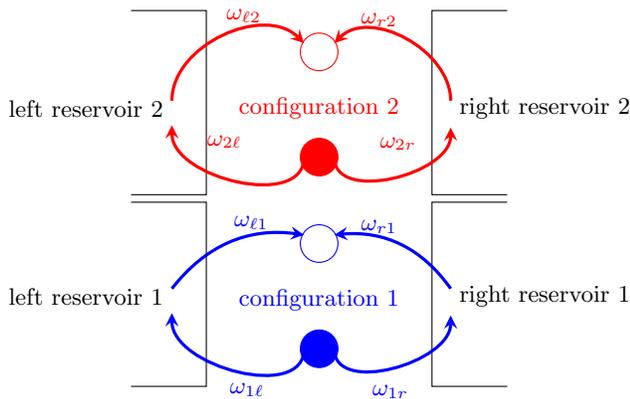

When in contact with a single reservoir, the probability distribution of occupation of the system will relax to an equilibrium distribution $p_{\mathrm{eq}}$. 
To complete our pump construction, we need to add the active, nonequilibrium ingredient. Since our intention is to provide an exact and explicit stochastic thermodynamic analysis for a periodically driven pump, we consider the simplest possible modulation. The connection of the system to the outside world is periodically alternated in a piece-wise constant way,  such that the corresponding equilibrium states are $p^{(1)}_{\mathrm{eq}}$ when $t\in[0,\tau/2], \mod \tau$, and $p^{(2)}_{\mathrm{eq}}$ for  $t\in[\tau/2,\tau], \mod \tau$, with $\tau$ the period. How this is achieved in detail is irrelevant for the subsequent analysis.  For the sake of clarity, we will focus on one possible implementation: the time-modulated two-state system is in contact with two ideal reservoirs at equilibrium.  As will be shown below, for a flux to exist the two contacts must be different from one another (``asymmetric coupling"). By switching the configuration of the set-up,  the equilibrium occupation $p_{\mathrm{eq}}$ of the system alternates between two different values. In the scenario in which the reservoirs are themselves not altered by the modulation, the difference in equilibrium occupation probabilities can be achieved by modulating the energy of the occupied state. The alternation is schematically reproduced in Fig.~\ref{fig:model}.

While in configuration $1$,  a particle may jump from the  system (when in the occupied state) to the left (right) reservoir with transition rate $\omega_{1\ell} (\omega_{1r})$, or from the left (right) reservoir to the system  (when in the empty state) with transition rate $\omega_{\ell 1} (\omega_{r 1})$. We also introduce the total rate from the system into
 either reservoir, $\omega_{10} = \omega_{1\ell} + \omega_{1 r}$, and the total rate from the reservoirs into the system,  $\omega_{01} = \omega_{\ell 1} + \omega_{r 1}$. Similarly, for the system in configuration $2$ the transition rates are denoted by $\omega_{2\ell}, \; \omega_{2 r}, \; \omega_{\ell 2}, \; \omega_{r 2}, \; \omega_{20}$ and $\omega_{02}$.  The corresponding equilibrium distributions are given by:
 \begin{eqnarray}
      p_{\mathrm{eq}}^{(1)}&=& \dfrac{\omega_{01}}{\omega_{01} +\omega_{10}},\\
     p_{\mathrm{eq}}^{(2)}& = & \dfrac{\omega_{02}}{\omega_{02} +\omega_{20}}.
       \end{eqnarray}

Since we assume that the system relaxes toward equilibrium when in a given configuration, detailed balance must be satisfied for the transitions between the system and each reservoir independently: 
  \begin{eqnarray}
      \omega_{1 \ell} p_{\mathrm{eq}}^{(1)} = \omega_{\ell 1} \left( 1 - p_{\mathrm{eq}}^{(1)} \right),&\; \;
      \omega_{1 r} p_{\mathrm{eq}}^{(1)}  = \omega_{r 1} \left( 1 - p_{\mathrm{eq}}^{(1)} \right),\\
       \omega_{2 \ell} p_{\mathrm{eq}}^{(2)} = \omega_{\ell 2} \left( 1 - p_{\mathrm{eq}}^{(2)} \right),& \; \;
      \omega_{2 r} p_{\mathrm{eq}}^{(2)} = \omega_{r 2} \left( 1 - p_{\mathrm{eq}}^{(2)} \right),
       \end{eqnarray}
    that is,
 \begin{eqnarray}
 \label{rr}
      p_{\mathrm{eq}}^{(1)}&=& \frac{\omega_{\ell 1}}{\omega_{\ell1} +\omega_{1\ell}}= \frac{\omega_{r 1}}{\omega_{r 1} +\omega_{1 r}},\\
      p_{\mathrm{eq}}^{(2)}&=& \frac{\omega_{\ell 2}}{\omega_{\ell 2}+\omega_{2 \ell}}= \frac{\omega_{r 2}}{\omega_{r 2} +\omega_{2 r}}.
 \label{rr2}
   \end{eqnarray}

   We end this section by introducing the ``reduced levels of occupancy", denoted by $\nu$, which will play a central role in the subsequent analysis:
  \begin{eqnarray}
    \nu_1&=&\frac{p_{\mathrm{eq}}^{(1)}}{1-p_{\mathrm{eq}}^{(1)}}
    = \frac{\omega_{\ell 1}}{\omega_{1\ell}} = \frac{\omega_{r 1}}{\omega_{1 r}}, \nonumber\\
  \nu_2&=&\frac{p_{\mathrm{eq}}^{(2)}}{1-p_{\mathrm{eq}}^{(2)}}
 = \frac{\omega_{\ell 2}}{\omega_{2\ell}} = \frac{\omega_{r 2}}{\omega_{2 r}}.
\label{eq:nu2}
\end{eqnarray}
  
  \section{Probability and Flux}
  \label{sec:prob}
  
 The time evolution of the probability vector ${\bf p}=\{1- p,p\}$  obeys a Markov equation
 \begin{equation}
 \dot{\bf p}={\bf M}{\bf p},
 \end{equation}
 where $\bf M$ is the time-periodic transition matrix. Its elements are specified in terms of the transition rates introduced above.  
 We are interested in the long-time solution of the Markov process. The probability $p(t)$ will then reach a ``steady'' time-periodic state with the same period as that of the modulation, $p(t)=p(t+\tau)$. 
 This function can be found as follows. The stochastic dynamics consists of a time-periodic alternation between two different  relaxations, 
  one toward $p_{\mathrm{eq}}^{(2)}$ and the other toward $p_{\mathrm{eq}}^{(1)}$ as the system switches periodically from configuration $2$ to $1$ at times equal to a multiple of $\tau$, and from $1$ back to $2$  at times $t=\tau/2$, mod $\tau$.  The unique steady state time-periodic solution is found by matching the end of this double relaxation after each period with the initial  value. The details of the calculation are given in the appendix. Denoting  the probability distributions when the modulation is in the first or second half of each period by $ p_1(t)$ and $p_2 (t)$,  one finds the following explicit results:

 \begin{widetext}
     \begin{eqnarray}
     \label{tps}
   p_1(t) &=&p_{\mathrm{eq}}^{(1)}
+  \dfrac{e^{-t (\omega_{0 1}+\omega_{10})} \left[1-e^{-\frac{1}{2} \tau (\omega_{0 2}+\omega_{2 0})}\right] 
   (\omega_{0 2} \omega_{1 0} - \omega_{0 1}\omega_{2 0})}
    {(\omega_{0 1}+\omega_{10 })(\omega_{0 2}+\omega_{2 0}) \left[1-e^{-\frac{1}{2} \tau
   (\omega_{0 1}+\omega_{0 2}+\omega_{10}+\omega_{2 0})} \right]},\\ 
p_2 (t) &=&p_{\mathrm{eq}}^{(2)}
+  \dfrac{e^{-(t-\tau) (\omega_{0 2}+\omega_{2 0})} \left[1 - e^{\frac{1}{2} \tau (\omega_{01}+\omega_{1 0})}\right] (\omega_{0 1} \omega_{2  0}-\omega_{0 2}\omega_{1 0})}{(\omega_{0 2}+\omega_{2 0 })(\omega_{0 1}+\omega_{10}) \left[1-e^{\frac{1}{2} \tau
   (\omega_{0 1}+\omega_{0 2}+\omega_{10}+\omega_{2 0})} \right]}. 
 \label{tps2}
  \end{eqnarray}
\end{widetext}
  
  As expected,  if the modulation is slow ($\tau\rightarrow \infty$), the probability distribution relaxes to the corresponding equilibrium distribution at the end of each half period. 
 In the  fast modulation  limit, on the other hand ($\tau \rightarrow 0$), the system freezes into the following nonequilibrium steady  state: 
  \begin{equation}
    p_1(t) \simeq p_2(t) \simeq \frac{\omega_{01}+\omega_{02}}{\omega_{01}+\omega_{10}+\omega_{02}+\omega_{20}}.
  \end{equation}
 This corresponds to the steady state for an unmodulated  system with effective transition rates $\omega_{01}+\omega_{02}$ and $\omega_{10}+\omega_{20}$.
   
\begin{figure}[htpb]
   \centering
   \includegraphics[width=1.0\linewidth]{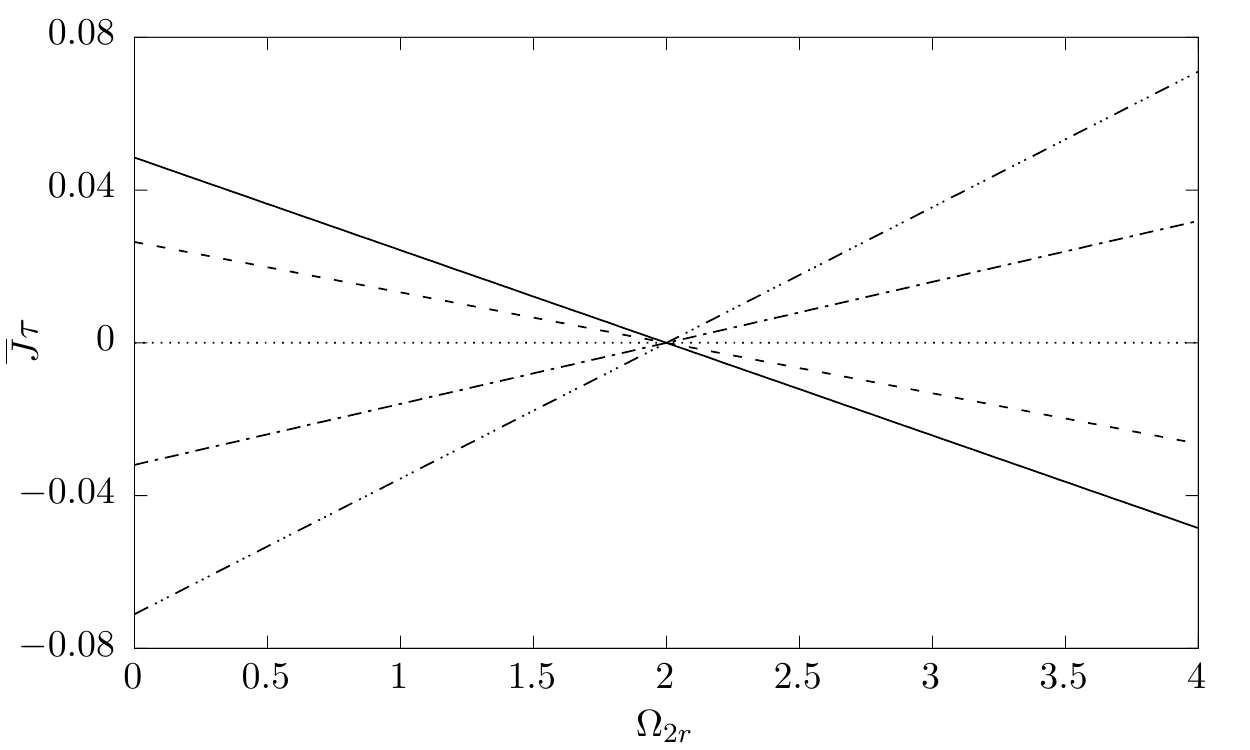}
   \caption{Average flux from the left reservoir (multiplied by the period of oscillation) as a function of the $\Omega_{2r}$. The parameters are $\Omega_{10}=1.5, \;\Omega_{1r}=1.0, \;\nu_1=0.8$ and $\Omega_{20}=3.0$. The different curves correspond, from top to bottom on the left-hand side of the figure, to the following values of   $\nu_2$: $0.2, 0.4, 0.8, 1.0$ and $1.2$.}
   \label{fig:reversal}
 \end{figure}

\begin{figure}[htpb]
   \centering
   \includegraphics[width=1.0\linewidth]{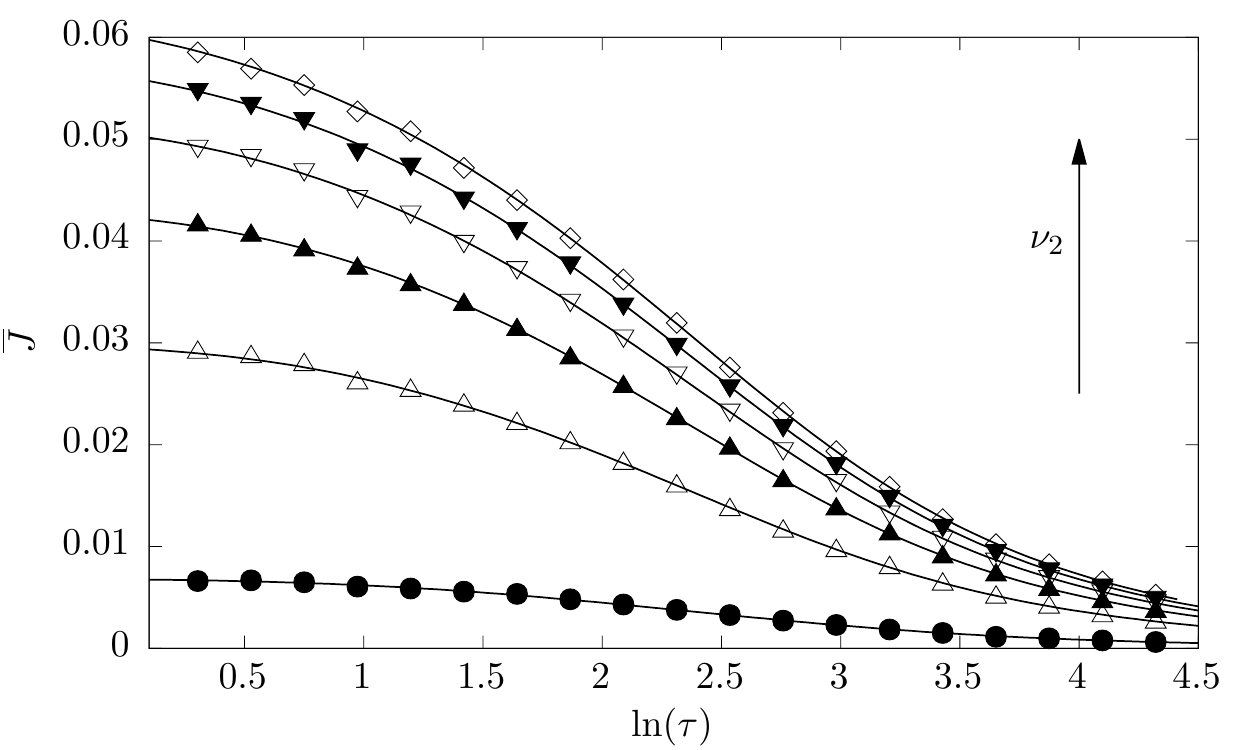}
   \caption{Net average flux from the left reservoir as a function of the period $\tau$ ($\ln$  scale). The other parameters are $\omega_{1\ell}=0.1, \;\omega_{1r}=0.2, \;\nu_1=0.1, \;\omega_{2\ell}=0.1$ and $\omega_{2r}=2.2$. The different curves correspond to the following values of $\nu_2$: $0.2, 0.7, 1.2, 1.7, 2.2$ and $2.7$. The symbols are the results of numerical simulations with $10$ samples of $10^5$ cycles each.}
   \label{fig:flux}
 \end{figure}

 We are now in a position to evaluate the net flux through the system. Since the system can at most carry a single particle, any net flux from one of the reservoirs to the system has to be compensated by a corresponding net flux out of the system into the other reservoir. Hence the system operates as a pump. The net average flux at time $t$ in each period coming from the left reservoir is given by:
 \begin{equation}
   J(t) = \left \{ \begin{array}{ll}
     \omega_{\ell 1} \left[1-p_1(t)\right ] - \omega_{1\ell} p_1(t) & \mathrm{for} \; 0 \leq t < \tau/2 \\
     \omega_{\ell 2} \left[1-p_2(t)\right] - \omega_{2\ell} p_2(t) & \mathrm{for} \; \tau/2 \leq t < \tau 
   \end{array}
   \right .
 \end{equation}
As we are focusing on the steady state time-periodic regime, the quantity of interest is the average of this quantity over one period:
 \begin{equation}
   \overline{J} = \frac{1}{\tau} \int_0^\tau J(t) dt.
 \end{equation}
In combination with Eq.~(\ref{eq:nu2}),  one finds:
\begin{widetext}
 \begin{equation} 
   \overline{J} = \frac{\left[e^{\frac{1}{2} (\nu_1+1) \tau \omega_{10}}-1\right] \left[e^{\frac{1}{2} (\nu_2+1) \tau \omega_{20}}-1\right]}{\tau \left[e^{\frac{1}{2} \tau (\nu_1 +1)\omega_{10}+\frac{1}{2} \tau (\nu_2 +1) \omega_{20}}-1\right]}\frac{\omega_{1\ell} \omega_{2 r}-\omega_{1 r} \omega_{2\ell}}{\omega_{10} \omega_{20}} \frac{\nu_1-\nu_2}{(\nu_1+1) (\nu_2+1)}.
   \label{eq:flowavg}
 \end{equation}
 \end{widetext}
This is the first main result of our paper, and we pause to make a few comments. Firstly, we note that the above expression incorporates the broken symmetries needed for the system to operate as a ratchet-like pump. Equilibrium corresponds to $\nu_1=\nu_2$, which is equivalent to $p_{\mathrm{eq}}^{(1)}=p_{\mathrm{eq}}^{(2)}$.
Despite being a trivial result, it is reassuring to see that the flux is zero in this case. Secondly, and again not very surprisingly, both states of the system must make contact with at least one reservoir. For example,  the flux vanishes if we set both $\omega_{2\ell}$ and $\omega_{2r}$ equal to zero. 
Thirdly, the sign of $\overline{J}$  changes  upon interchanging the left and right reservoirs. An interesting consequence is that no flux exists when the system obeys the left-right symmetry  $\omega_{1r} \omega_{2\ell}=\omega_{1\ell} \omega_{2r}$. 
Fourthly, the signs of the products of the differences $\nu_1 - \nu_2$ and $\omega_{1\ell}\omega_{2r}-\omega_{1r}\omega_{2\ell}$ determine the direction of the flow, with equilibrium and the symmetric situation being points of flux reversal. 
We illustrate this phenomenon in  Fig.~\ref{fig:reversal}. Introducing the variables $
   \Omega_{\alpha} = \omega_{\alpha} \tau$, we note that 
   flux reversal occurs when $\Omega_{2r} = \omega_{2\ell}{\omega_{1r}}/{\omega_{1\ell}}= \Omega_{1r} {\Omega_{20}}/{\Omega_{10}}  = 2$. 
Fifthly, we mention the limits of slow oscillation ($\tau\rightarrow \infty$) and  fast oscillation ($\tau\rightarrow 0$). In the former case, the exponentials in  numerator and denominator cancel and the flux decays as $1/\tau$: 
 \begin{equation}
   \lim_{\tau\rightarrow \infty} \overline{J} \sim  \frac{1}{\tau}\frac{(\nu_1-\nu_2)(\omega_{1 \ell} \omega_{2r}-\omega_{1 r} \omega_{2\ell})}{(\nu_1+1) (\nu_2+1) \omega_{10} \omega_{20}}.
   \label{eq:flowavginf}
 \end{equation}
 For fast oscillations the average flux tends to a nonzero constant value:
 \begin{equation}
   \lim_{\tau\rightarrow 0} \overline{J} = \frac{(\nu_1-\nu_2) (\omega_{1\ell} \omega_{2r}-\omega_{1r} \omega_{2\ell})}{2 \left[(\nu_1+1) \omega_{10}+(\nu_2+1)
   \omega_{20}\right]}.
   \label{eq:flowavg0}
 \end{equation}
 Finally, we notice that the average flux $\overline{J}$ is a monotonic function of the period. In fact, its absolute value decreases as the period of oscillation increases. This monotonic decay can be observed in Fig.~\ref{fig:flux}. This figure also shows the increase in the flux as we move away from equilibrium, that is, as $\nu_2 (>\nu_1)$ increases. The figure illustrates the perfect agreement of our exact expression for the average flux with numerical simulations. 
 
 \section{Entropy Production}
 
 \label{sec:entropy_production}
 
We start from the general definition for the rate of entropy production of a Markov process characterized by transition rates $M_{ij}$ between states $i$ and $j$. Following stochastic thermodynamics \cite{prigogine,groot,tome,vandenbroeck2016}, it is given by:
\begin{equation}
   \dot{S} = k_B \sum_{ij} (M_{ij}P_j-M_{ji}P_i) \ln\frac{M_{ij}P_j}{M_{ji}P_i}. 
  \end{equation}
  Here $P_i$ $(P_j)$ is the probability that the system is in state $i$ (state $j$).
Applied to our model, we get the  following expression in terms of the probabilities $p_i$ that the system is in the occupied state while in configuration $i$:
\begin{widetext}
\begin{equation}
   \dot{S}(t) = k_B \left \{ \begin{array}{ll}
     \sum_{k=\ell, r} \left [\omega_{1k} p_1(t) - \omega_{k1}\left(1- p_1(t)\right)\right ] \ln\dfrac{\omega_{1k} p_1(t)}{\omega_{k1}\left(1- p_1(t)\right)} & \mathrm{for} \; 0 \leq t < \tau/2 \\
     \sum_{k=\ell, r} \left [\omega_{2k} p_2(t) - \omega_{k2}\left(1- p_2(t)\right)\right ] \ln\dfrac{\omega_{2k} p_2(t) }{\omega_{k2} \left(1-p_2(t)\right)} & \mathrm{for} \; \tau/2 \leq t < \tau.
   \end{array}
   \right.
 \end{equation}
 \end{widetext}
 
 To find an appropriate ``steady state'' expression characterizing the periodically operating pump, 
 we need to perform an average over one cycle:
 \begin{equation}
  \overline{\dot{S}_i}= \frac{1}{\tau} \int_0^\tau \dot{S}_i(t) dt.
 \end{equation}
 After a strenuous calculation, one obtains the following simple and revealing expression, which is our second major result:
\begin{widetext}
\begin{equation}
 \overline{\dot{S}} = k_B \frac{\left[e^{\frac{1}{2} (\nu_1+1) \tau \omega_{10}}-1\right] \left[e^{\frac{1}{2} (\nu_2+1) \tau \omega_{20}}-1\right]}{\tau \left[e^{\frac{1}{2} (\nu_1+1) \tau \omega_{10}+\frac{1}{2} (\nu_2+1) \tau \omega_{20}}-1\right]}\frac{(\nu_1-\nu_2)}{(\nu_1+1) (\nu_2+1)} \ln \frac{\nu_1}{\nu_2}.
  \label{eq:entprod}
\end{equation}
\end{widetext}
In the next section we will recover this result via a less strenuous approach using stochastic thermodynamics for a particular case. 

We again pause to make several comments. First, the entropy production is positive, as it should be. It vanishes and only vanishes at equilibrium, $\nu_1=\nu_2$, as it should.  Second, by combination with the expression for the flux, cf. Eq.~(\ref{eq:flowavg}), the entropy production can be written as a flux-times-force expression, familiar from irreversible thermodynamics:
 \begin{equation}\label{ejx}
   \overline{\dot{S}}= \overline{J}X,
 \end{equation}
with the following expression for the thermodynamic force:
 \begin{equation}
 \label{force}
 X= k_B \frac{\omega_{10}\omega_{20}}{\omega_{1\ell} \omega_{2r}-\omega_{1r} \omega_{2\ell}}  \ln \frac{\nu_1}{\nu_2}.
 \end{equation}

\begin{figure}[htpb]
   \centering
   \includegraphics[width=1.0\linewidth]{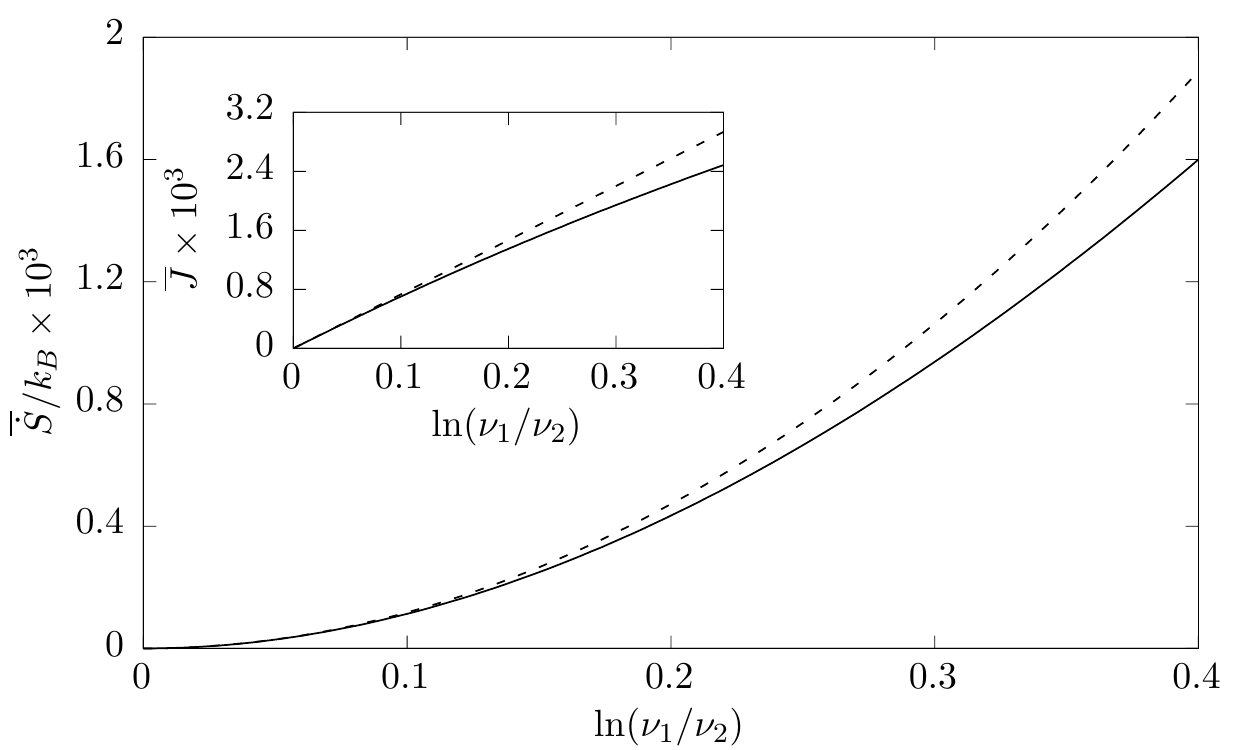}
   \caption{Comparison of the exact (full line) and Onsager approximation (dashed line) expressions for the average entropy production (main panel) and flux (inset) per cycle for small forces. The parameter values are $\omega_{1\ell}=0.1, \;\omega_{1 r}=0.2, \;\nu_1=0.1, \;\omega_{2\ell}=2.1$ and $\omega_{2 r}=0.1$. }
   \label{fig:entprodonsager}
 \end{figure}

Recalling that $\nu=p_{\mathrm{eq}}/(1-p_{\mathrm{eq}})$ and that $p_{\mathrm{eq}}$ is the equilibrium probability for an occupied system, this expression for $X$ reproduces the intuitive observation that its amplitude depends on the (logarithmic) difference between  occupation in both configurations. The sign of the force is, however, also determined by
the  balance of rates (cf.\ denominator $\omega_{1\ell} \omega_{2r}-\omega_{1r} \omega_{2\ell}$).
Third, the corresponding Onsager coefficient that describes the linear response regime is then found by evaluating the flux $\overline{J}$ in the limit of small force, $X\rightarrow 0$, or equivalently, in the equilibrium limit $\nu_1\rightarrow \nu_2\equiv \nu$: 
\begin{equation}
  \overline{J} \sim LX.
 \end{equation}
One finds: 
\begin{widetext}
\begin{equation}
L=\frac{\nu}{k_B (1+\nu)^2}\frac{\left[e^{\frac{1}{2} (\nu+1) \tau \omega_{10}}-1\right] \left[e^{\frac{1}{2} (\nu+1) \tau \omega_{20}}-1\right]}{\tau \left[e^{\frac{1}{2}(\nu +1)\tau(\omega_{10}+ \omega_{20})}-1\right]}  \left(\frac{\omega_{1\ell} \omega_{2 r}-\omega_{1r} \omega_{2\ell}}{\omega_{10}\omega_{20}}\right)^2.
 \end{equation}
 \end{widetext}
 As expected, the Onsager coefficient is invariant upon interchange of the left and right reservoirs. It is always positive, reflecting that current $\overline{J}$ flows in the direction of the force $X$.  Fourth, we note that the results for the flux and entropy production are exact, and valid far from equilibrium. In Fig.~\ref{fig:entprodonsager}, we show how the exact entropy production expression Eq.~(\ref{eq:entprod}) deviates from its  near equilibrium expression (main panel): 
\begin{equation}\label{epe}
  \overline{\dot{S}_i}=L X^2,
\end{equation}
and how the linear flux-versus-force relation (inset) breaks down. The figure shows that the entropy production is greater when the system is farther from equilibrium.
The graphs are plotted  as functions of  $\ln \nu_1/\nu_2$, used here as  a measure of the distance from equilibrium.

\begin{figure}[htpb]
   \centering
   \includegraphics[width=1.0\linewidth]{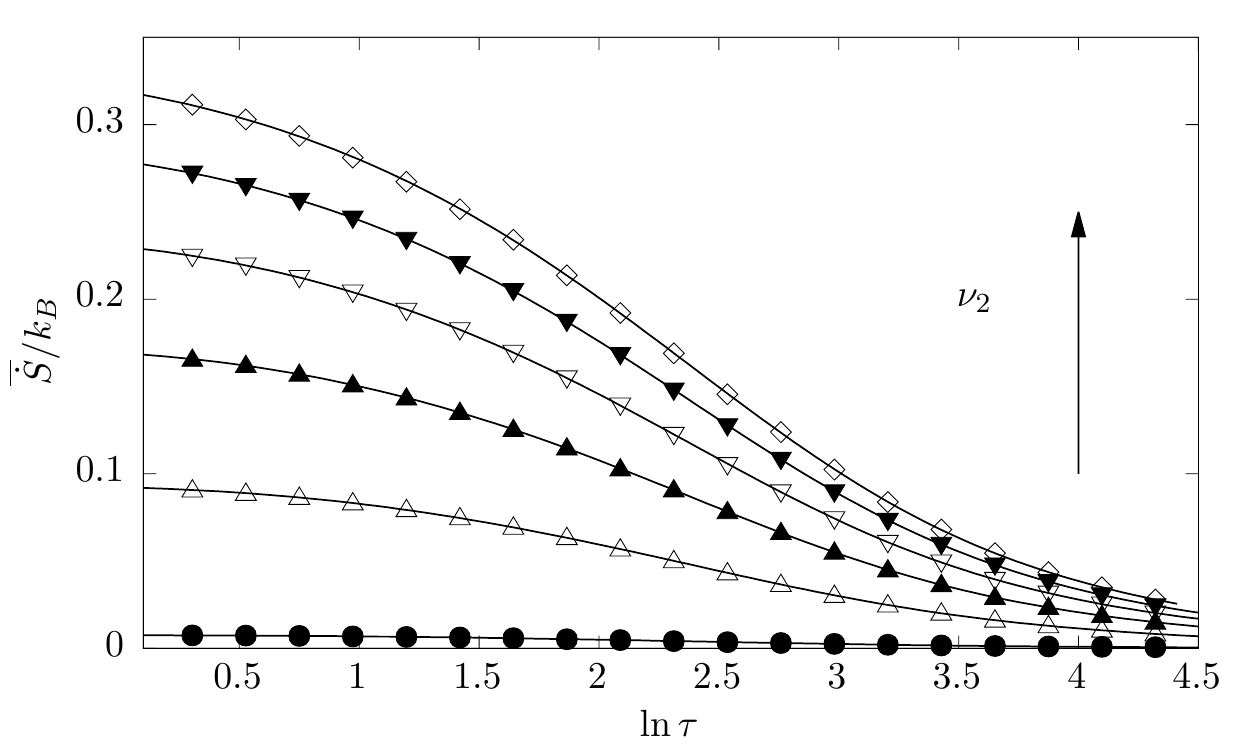}
   \caption{Average entropy production per cycle as a function of the period). The parameters are $\omega_{1\ell}=0.1, \;\omega_{1 r}=0.2, \;\nu_1=0.1 \;\omega_{2\ell}=2.1$ and $\omega_{2 r}=0.1$. The different curves correspond to different values of $\nu_2$ -- from bottom to top: 0.2, 0.7, 1.2, 1.7, 2.2 and 2.7. The symbols are the results of numerical simulations with 100 samples of 10,000 cycles each.}
   \label{fig:entprod}
 \end{figure}

 Finally, we carried out extensive numerical simulations and found perfect agreement with the above analytic result for the entropy production, cf.  Fig.~\ref{fig:entprod}.

 \section{Work, Heat  and Efficiency}
 So far we have made no reference to the concept of energy. In this section, we  consider a scenario that connects the above construction to a  thermo-chemical pump engine \cite{esposito2009,esposito2010a,esposito2010b}. This will  allow us to ask the standard thermodynamic questions about work, heat and efficiency.
 To simplify matters, we take the system to be in contact with a single reservoir in each of its two configurations, say the left reservoir in configuration $1$ and the right reservoir in configuration $2$. Mathematically, this is achieved by taking the limits:
 \begin{equation}\label{limit} 
 \omega_{2\ell}\rightarrow 0,\;\;\;\omega_{\ell 2}\rightarrow 0,\;\;\;\omega_{1 r}\rightarrow 0,\;\;\;\omega_{r 1}\rightarrow 0. 
 \end{equation}
 Next, we attribute the energies $\epsilon_1$ and $\epsilon_2$ to the system when occupied by a particle in configurations $1$ and $2$, respectively. This implies that upon the transitions from configuration $1$ to $2$ and back, while the system contains a particle, an energy equal to $\epsilon_2-\epsilon_1$ and $\epsilon_1-\epsilon_2$ has to be provided via an outside source, which we take to be a dissipationless work source.  Averaged over one period, and recalling that the flux $\overline{J}$ is measured from the left reservoir into the system, we conclude that the work  $\overline{W}$ on the system is given by 
\begin{equation}
  \overline{W}=\overline{J}(\epsilon_2-\epsilon_1). 
  \label{eq:work}
\end{equation}
Furthermore, the left and right reservoirs are characterized by chemical potentials and temperatures equal to $\mu_{\ell}, T_{\ell}$  and $\mu_{r},T_{r}$, respectively. The particle pump thus produces a period-averaged amount of chemical work $\overline{W}_{\mathrm{chem}}$ from the system into the reservoirs, and heat currents  $\overline{Q}_{\mathrm{l}}$ and $\overline{Q}_{\mathrm{r}}$ from the left and right reservoirs into the system, 
 given by: 
 \begin{eqnarray}
   \overline{W}_{\mathrm{chem}}&=&\overline{J}(\mu_{r}-\mu_{\ell}),\nonumber\\
    \overline{Q}_{\ell}&=&\overline{J}(\epsilon_1-\mu_{\ell}),\\
    \overline{Q}_{r}&=&-\overline{J}(\epsilon_2-\mu_{r}).\nonumber
   \label{eq:heats}
 \end{eqnarray}
Taking into account that the system returns to the same (statistical) state after each period, the first law
requires that the sum of all average energy contributions vanishes. Noting that the chemical work is work provided to the reservoirs, we get the following energy balance equation:
 \begin{equation}
\overline{W}+\overline{Q}_{\ell}+\overline{Q}_{r}= \overline{W}_{\mathrm{chem}}.
 \end{equation}
The second law is recovered by noting that  the heat fluxes are responsible for the  entropy production, and hence
 \begin{equation}
   \overline{\dot{S}}=-\frac{\overline{Q}_{\ell}}{T_{\ell}}-\frac{\overline{Q}_{r}}{T_{r}}.
 \end{equation}
 By identification with $ \overline{\dot{S}}= \overline{J}X$, cf. Eq.~(\ref{ejx}), one thus obtains the following expression for the thermodynamic force $X$:
 \begin{equation}\label{force2}
 X=\frac{\epsilon_2-\mu_{r}}{T_{r}}-\frac{\epsilon_1-\mu_{\ell}}{T_{\ell}}.
 \end{equation}
The consistency of these expressions with the previous results for entropy production and thermodynamic force, cf. Eqs.~(\ref{eq:entprod}),  (\ref{ejx}), and (\ref{force}),  comes through the 
 explicit identification of $p_{\mathrm{eq}}$  for  a system in contact with a heat-particle reservoir $\mu, T$. There are only two energy states for the system, namely, energy equal to $0$ (empty) and equal to $\epsilon$ (occupied). The probability for the occupied state is given by the Fermi function:
 \begin{equation}
 p_{\mathrm{eq}}=\frac {1}{e^{\beta(\epsilon-\mu)}+1}.
 \end{equation}
 This result fixes the ratio of the exchange rates with each of the reservoirs, cf. Eqs.~(\ref{rr}) and (\ref{eq:nu2}). It suffices to verify that, with this prescription, Eq.~(\ref{force2}) indeed reduces to Eq.~(\ref{force}). The equivalence is established by remembering the limits Eq.~(\ref{limit}), implying that  Eq.~(\ref{force}) simplifies to $X=k_B \ln \nu_1/\nu_2$, together with:
\begin{equation}
 \ln \nu=\ln p_{\mathrm{eq}}/(1-p_{\mathrm{eq}})=-\beta (\epsilon-\mu).
 \label{eq:lognu}
\end{equation} 
 
 To discuss the issue of efficiency, we focus on the case of a thermal engine, with the left reservoir playing the role of the hot, heat providing entity ($\overline{Q}_{\ell}>0$, ${T_{\ell}}>{T_{r}}$). The output is the net work, i.e., the chemical work minus the input work. The efficiency $\eta$ is thus given by:
 \begin{equation}
 \eta=\frac{\overline{W}_{\mathrm{chem}}-\overline{W}}{\overline{Q}_{\ell}}=1+\frac{\overline{Q}_{r}}{\overline{Q}_{\ell}}=1-\frac{T_{r}}{T_{\ell}}
 -\frac{T_{r}  \overline{\dot{S}_i}}{\overline{Q}_{\ell}}\leq 1-\frac{T_{r}}{T_{\ell}}.
 \label{eq:etadef}
 \end{equation}
 
 We thus have explicit analytic expressions for the power, efficiency and dissipation valid at any distance away from equilibrium. Actually, combining the expression for the efficiency, cf. Eq.~(\ref{eq:etadef}), with those for the heat currents, cf. Eq.~(\ref{eq:heats}), one  concludes that the efficiency can be rewritten as follows:
 \begin{equation}
   \eta = 1 - \frac{\epsilon_2 - \mu_{r}}{\epsilon_1 - \mu_{\ell}}.
   \label{eq:eta7}
 \end{equation}
 The efficiency is thus fully determined by the choice of energy levels.
 At first glance, this observation is surprising  because none of the parameters that are related to the dissipation appear explicitly in Eq.~(\ref{eq:eta7}), i.e., the rate of modulation,  the temperatures and the rate of entropy production. 
However, one must remember that the energies are linked to the variable $\nu$ and temperature via Eq. (\ref{eq:lognu}). In fact, one immediately verifies that Carnot efficiency is recovered if one specifies that the system operates under equilibrium conditions, $X=0$ or $({\epsilon_2-\mu_{r}}){T_{\ell}}=({\epsilon_1-\mu_{\ell}}){T_{r}}$, cf. Eq.~(\ref{force2}).
Note also that,  as we move further away from equilibrium, the efficiency decreases linearly with $\ln (\nu_1/\nu_2)$ and eventually becomes negative. The latter regime corresponds to a dud engine, as it just dissipates while failing to deliver any work at all.
  
\begin{figure}[htpb]
   \centering
   \includegraphics[width=0.8\linewidth]{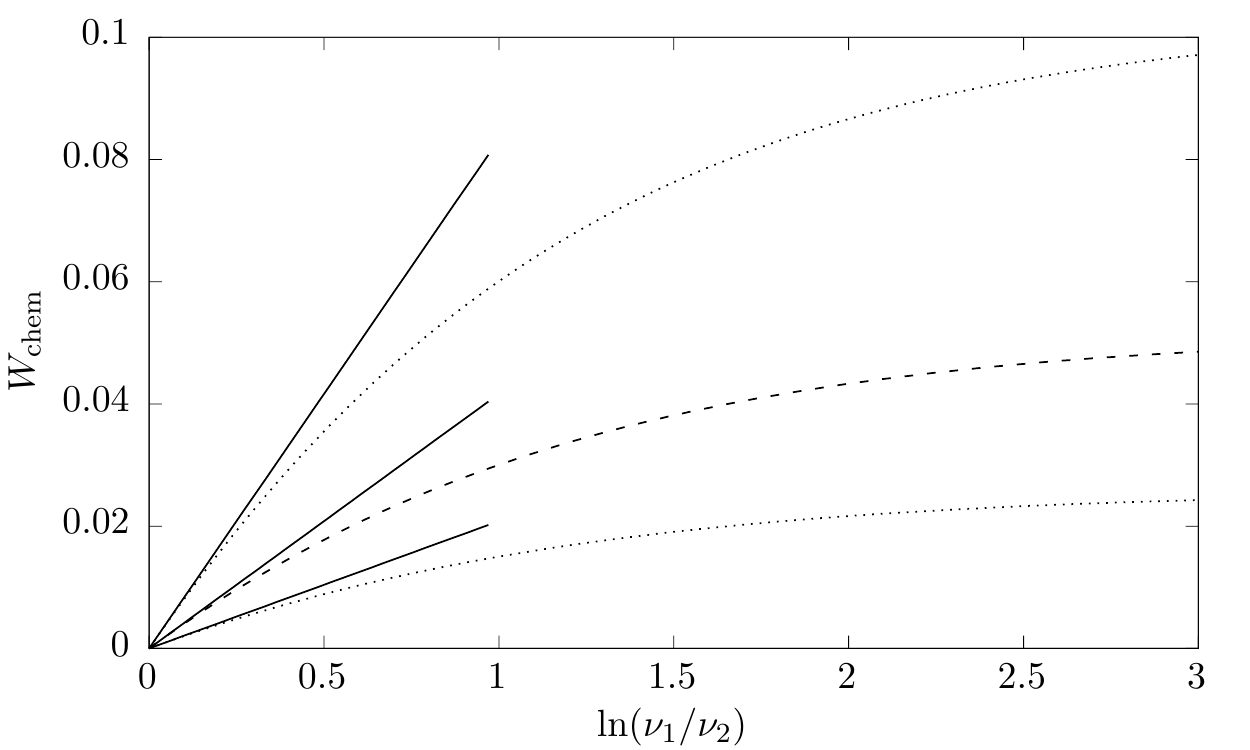}
   \caption{Chemical work as a function of the thermodynamic force. Parameter values: $\epsilon_1=1, \;\mu_{\ell}=0, \; \beta_{\ell}=1, \; \beta_{r}=2, \; \omega_{1\ell}=1, \omega_{1 r}=2$ and $\tau=1$. The three curves correspond to the following values of $\mu_{r}: 1, \; 1/2$, and $1/4$ (from top to bottom). The straight lines correspond to the linear response approximation.}
   \label{fig:wchem}
 \end{figure}

We close this section with two  illustrative plots. In Fig.~\ref{fig:wchem} we show how the chemical work increases as we move away from equilibrium. Near equilibrium, the linear response approximation agrees with the exact solution, but  it overestimates the chemical work, as the latter saturates  in tune with the particle flux.
  
\begin{figure}[htpb]
   \centering
   \includegraphics[width=0.8\linewidth]{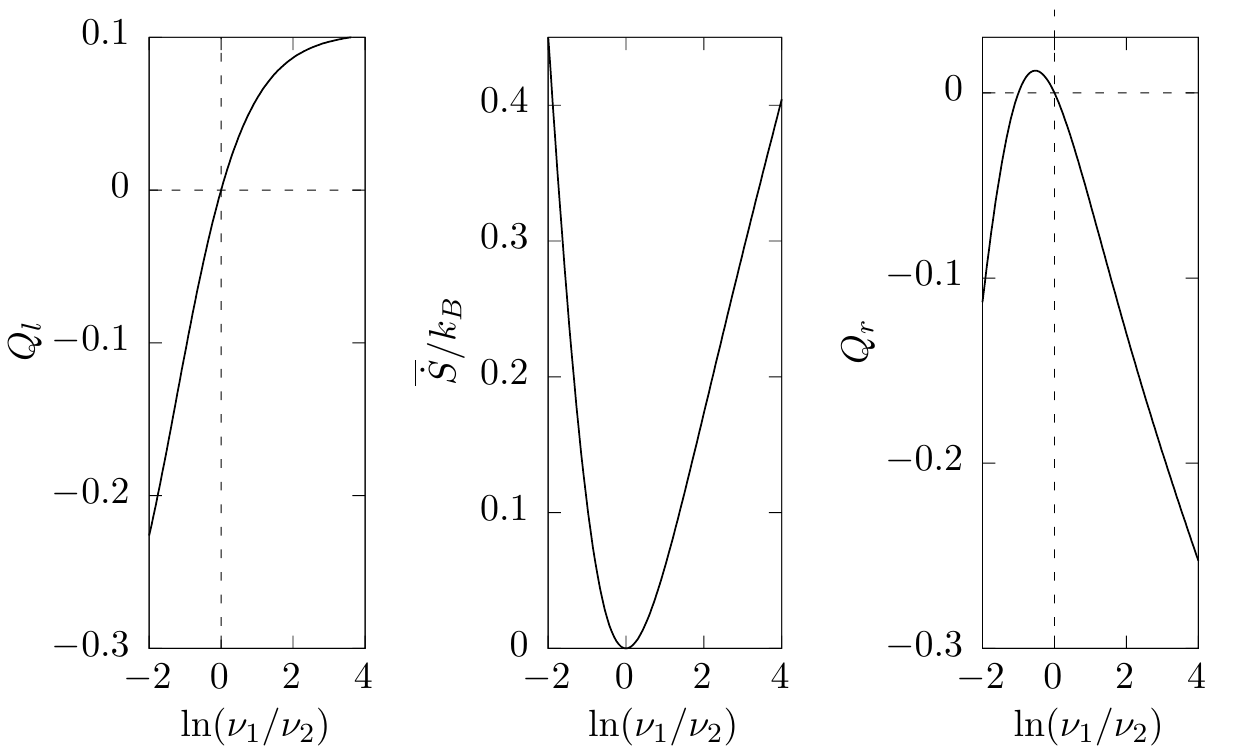}
   \caption{Heat current from the left reservoir (left panel), entropy production (middle panel) and heat current from the right reservoir (right panel) as a function of the thermodynamic force.  Same parameter values as in the previous figure.}
   \label{fig:heats}
 \end{figure}

In Fig.~\ref{fig:heats} we reproduce the heat current from the left and right reservoirs, together with the entropy production.  The direction of both heat currents  reverses at equilibrium, while the entropy production reaches its minimum value (zero). 
 Note that the  plotted curves are, for all three quantities, independent of $\mu_{r}$. This can be understood from the fact that changing $\mu_{r}$ implies a corresponding change of $\epsilon_2$ such as to keep a  fixed value of $\nu_2$ (with all other parameters also being held constant). Concomitantly, the flux is the same regardless of the value of this chemical potential. 

  \section{Perspectives}
 We have introduced a simple model of a periodically driven single particle pump. It is exactly solvable and amenable to a full and detailed stochastic thermodynamic analysis. It will allow to verify and test other predictions as they arise  from stochastic thermodynamics. One example is the recently derived thermodynamic uncertainty relation for periodically driven systems \cite{proesmans2017}.  The calculations presented here can also be repeated for a model with  $3$ instead of $2$ configurations. Such a construction allows one to break the strong coupling constraint which requires the energy and particle flows to be  proportional to each other. This will make it possible to study the symmetry properties of both the linear and nonlinear Onsager coefficients \cite{andrieux}.

 \begin{acknowledgments}
   AR thanks the CNPq (Grant No. 307931/2014-5) for financial support. KL acknowledges the support of the US Office of Naval Research (ONR) under Grant No. N00014-13-1-0205.
 \end{acknowledgments}
 \appendix*
 \section{Steady state distribution}
   We first focus our attention on the general relaxation dynamics when the system is in configuration 1. The transition matrix for this system is
  \begin{equation}
    M = \left( \begin{array}{cc}
      -\omega_{01} & \omega_{10} \\
      \omega_{01} & -\omega_{10} 
    \end{array} \right).
  \end{equation}
  This matrix has two eigenvalues: $\lambda^{(1)}_{\mathrm{eq}}=0$ (corresponding to the equilibrium state), and $\lambda^{(1)}_- = -\omega_{10} - \omega_{01}$ (which governs the decay to equilibrium). The corresponding eigenvectors are
  \begin{equation}
    |\Psi^{(1)}_{\mathrm{eq}}\rangle = \left( \begin{array}{c}
      \alpha_{\mathrm{eq}} \\
      \frac{\omega_{01}}{\omega_{10}} \alpha_{\mathrm{eq}}
    \end{array} \right), 
    \quad \mathrm{and} \quad
    |\Psi^{(1)}_- \rangle = \left( \begin{array}{c}
      \alpha_{-} \\
      \frac{\omega_{01}}{\omega_{01}+\omega_{10}} \alpha_{-}
    \end{array} \right), 
  \end{equation}
  where $\alpha_{\mathrm{eq}}$ and $\alpha_-$ are constants. Defining the inner product of two vectors 
  \begin{equation}
    |f\rangle = \left( \begin{array}{c}
      f_1\\
      f_2
    \end{array} \right), 
    \quad \mathrm{and} \quad
    |g\rangle = \left( \begin{array}{c}
      g_1\\
      g_2
    \end{array} \right), 
  \end{equation}
  as
  \begin{equation}
    \langle f | g \rangle = \frac{f_1 g_1}{\alpha_{\mathrm{eq}}} +\frac{f_2 g_2}{\frac{\omega_{01}}{\omega_{10}}\alpha_{\mathrm{eq}}},
  \end{equation}
  and imposing the normalization condition $\langle \Psi^{(1)}_{\mathrm{eq}} | \Psi^{(1)}_{\mathrm{eq}} \rangle =1,$ we have that 
  \begin{equation}
    \alpha_{\mathrm{eq}} = \frac{\omega_{10}}{\omega_{01}+\omega_{10}},
  \end{equation}
  so that,
  \begin{equation}
    |\Psi^{(1)}_{\mathrm{eq}}\rangle = \left( \begin{array}{c}
      \frac{\omega_{10}}{\omega_{01}+\omega_{10}}\\
      \frac{\omega_{01}}{\omega_{01}+\omega_{10}}
    \end{array} \right), 
    \quad \mathrm{and} \quad
    \langle \Psi^{(1)}_{\mathrm{eq}}| = \left( \begin{array}{cc}
      1 & 1
    \end{array} \right).
  \end{equation}
  Analogously, the normalization condition for the eigenvector $| \Psi^{(1)}_- \rangle$ leads to
  \begin{equation}
    |\Psi^{(1)}_{-}\rangle = \left( \begin{array}{c}
    \frac{\sqrt{\omega_{01} \omega_{10}}}{\omega_{01}+\omega_{10}}\\
    -\frac{\sqrt{\omega_{01} \omega_{10}}}{\omega_{01}+\omega_{10}}
    \end{array} \right),
    \quad \mathrm{and} \quad
    \langle \Psi^{(1)}_{-}| = \left( \begin{array}{cc}
      \sqrt{\frac{\omega_{01}}{\omega_{10}}} & -\frac{\omega_{10}}{\sqrt{\omega_{01} \omega_{10}}}
    \end{array} \right).
  \end{equation}
  Therefore, if the system is in state
  \begin{equation}
    |P_0\rangle = \left( \begin{array}{c}
      1 - p_0 \\
      p_0
    \end{array} \right),
  \end{equation}
at time $t=0$, it will evolve to equilibrium so that
  \begin{equation}
    |P_1(t)\rangle = \left( \begin{array}{c}
      1 - p_1(t) \\
      p_1(t)
    \end{array} \right),
  \end{equation}
  with
  \begin{equation}
    p_1(t) = p_{\mathrm{eq}}^{(1)} + \left( p_0 - p_{\mathrm{eq}}^{(1)}  \right) e^{\lambda_-^{(1)}t}, 
    \label{eq:p1t}
  \end{equation}
  where $p_{\mathrm{eq}}^{(1)} = \omega_{01}/(\omega_{01}+\omega_{10})$.

  Now, for the two-configuration system, Eq.~(\ref{eq:p1t}) still governs the time-evolution of the system while it is in configuration 1. Hence, if we start our clock when the system goes to configuration 1, Eq.~(\ref{eq:p1t}) will hold up to $\tau/2$ (when the system jumps to configuration 2). In the following half period, the time-evolution will be governed by the configuration 2 dynamics, that is,
  \begin{equation}
    p_2(t) = p_{\mathrm{eq}}^{(2)} + \left[ p_1(\tau/2) - p_{\mathrm{eq}}^{(2)}  \right] e^{\lambda_-^{(2)}(t-\tau/2)}, 
    \label{eq:p2t}
  \end{equation}
  where $p_{\mathrm{eq}}^{(2)} = \omega_{02}/(\omega_{02}+\omega_{20})$ and $\lambda^{(2)}_- = -\omega_{20} - \omega_{02}$.

  As stated above, we are interested in the steady state. Therefore, after a complete cycle, the system must return to the beginning state. Consequently, $p_2(\tau) = p_0$, which leads to
  \begin{equation}
    p_0 = \frac{e^{\frac{\lambda_-^{(2)} \tau}{2}} \left[p_{\mathrm{eq}}^{(1)} \left(e^{\frac{\lambda_-^{(1)} \tau}{2}}-1\right)+p_{\mathrm{eq}}^{(2)}\right]-p_{\mathrm{eq}}^{(2)}}{e^{\frac{1}{2} \tau (\lambda_-^{(1)}+\lambda_-^{(2)})}-1}.
  \end{equation}
  Substituting this result in the expressions for $p_1(t)$ and $p_2(t)$ we obtain the result of Eqs.~(\ref{tps}) and~(\ref{tps2}) from the main text.


\begin{thebibliography}{10}
\bibitem{schmiedl} S. Schmiedl and U Seifert, Europhys. Lett. \textbf{81}, 20003 (2008).
\bibitem{izumida1} Y. Izumida and K. Okuda,  Phys. Rev. E \textbf{80}, 021121 (2009).
\bibitem{izumida2} Y. Izumida and K. Okuda, Eur. Phys. J. B \textbf{77}, 499 (2010).
\bibitem{esposito} M. Esposito, R. Kawai, K. Lindenberg and C. Van den Broeck, Phys. Rev. E \textbf{81}, 041106 (2010).
\bibitem{izumida3} Y. Izumida and K. Okuda, Europhys. Lett. \textbf{97}, 10004 (2012).
\bibitem{izumida4} Y. Izumida and K. Okuda, New J. Phys. \textbf{17}, 085011 (2015).
\bibitem{brandner1} K. Brandner, K. Saito and U. Seifert, Phys. Rev. X \textbf{5}, 031019 (2015).
\bibitem{proesmans1} K. Proesmans and C. Van den Broeck, Phys. Rev. Lett. \textbf{115}, 090601 (2015).
 \bibitem{proesmans2} K. Proesmans, B. Cleuren and C. Van den Broeck, J. Stat. Mech. 023202 (2016).
 \bibitem{benenti} G. Benenti, K. Saito, and G. Casati, Phys. Rev. Lett. \textbf{106}, 230602 (2011).
 \bibitem{brandner2} K. Brandner, K. Saito, and U. Seifert, Phys. Rev. Lett. \textbf{110}, 070603 (2013). 
 \bibitem{proesmans3} K. Proesmans, B. Cleuren, and C. Van den Broeck, Phys. Rev. Lett. \textbf{116}, 220601 (2016).
 \bibitem{rosas16} A. Rosas, C. Van den Broeck, and K. Lindenberg, Phys. Rev. E. \textbf{94}, 052129 (2016). 
 \bibitem{rosas16b} A. Rosas, C. Van den Broeck, and K. Lindenberg, J. Phys. A: Math. Theor. \textbf{49}, 484001 (2016).
  \bibitem{proesmann4} K. Proesmann, Y. Dreher, M. Gavrilov, J. Bechhoefer and C. Van den Broeck, Phys. Rev. X \textbf{6}, 041010 (2016).
  \bibitem{proesmann5} K. Proesmann and C. Van den Broeck, ``The underdamped Brownian duet and stochastic linear irreversible thermodynamics'', to appear in Chaos.
 \bibitem{prigogine} I. Prigogine, \textit{Introduction to Thermodynamics of Irreversible Processes} (Interscience Publishers, New York, 1961).
 \bibitem{groot} S. R. De Groot and P. Mazur, \textit{Non-equilibrium Thermodynamics} pp. 61-83 (North-Holland Publ. Co., Amsterdam, 1962).
\bibitem{tome} T. Tom\'e and M. J. de Oliveira, Phys. Rev. E \textbf{91}, 042140 (2015).
\bibitem{vandenbroeck2016} C. Van den Broeck, S. Sasa and U. Seifert, New J. of Phys. \textbf{18}, 020401 (2016). 
\bibitem{esposito2009} M. Esposito, K. Lindenberg, and C. Van den Broeck, 
EPL \textbf{85}, 60010, 1-5 (2009).
\bibitem{esposito2010a}  M. Esposito, R. Kawai, K. Lindenberg, and C. Van den Broeck, 
EPL \textbf{89}, 20003 (2010).
\bibitem{esposito2010b} M. Esposito, R. Kawai, K. Lindenberg, and C. Van den Broeck, 
Phys. Rev. E \textbf{81}, 041106 (2010).
\bibitem{proesmans2017} K. Proesmans and C. Van den Broeck, 
EPL \textbf{119}, 20001 (2017). 
\bibitem{andrieux} D. Andrieux and P. Gaspard, J. Chem. Phys. \textbf{121}, 6167 (2004).
\end{thebibliography}
\end{document}